# Time-of-flight imaging of invisibility cloaks

**Jad C. Halimeh[1] and Martin Wegener[2]**

[1]*Physics Department, Arnold Sommerfeld Center for Theoretical Physics, and Center for NanoScience, Ludwig-Maximilians-Universität München, D-80333 München, Germany*
[2]*Institut für Angewandte Physik, DFG-Center for Functional Nanostructures (CFN), and Institut für Nanotechnologie, Karlsruhe Institute of Technology (KIT), D-76128 Karlsruhe, Germany*
[*]*Jad.Halimeh@physik.lmu.de*

**Abstract:** As invisibility cloaking has recently become experimental reality, it is interesting to explore ways to reveal remaining imperfections. In essence, the idea of most invisibility cloaks is to recover the optical path lengths without an object (to be made invisible) by a suitable arrangement around that object. Optical path length is proportional to the time of flight of a light ray or to the optical phase accumulated by a light wave. Thus, time-of-flight images provide a direct and intuitive tool for probing imperfections. Indeed, recent phase-sensitive experiments on the carpet cloak have already made early steps in this direction. In the macroscopic world, time-of-flight images could be measured directly by light detection and ranging (LIDAR). Here, we show calculated time-of-flight images of the conformal Gaussian carpet cloak, the conformal grating cloak, the cylindrical free-space cloak, and of the invisible sphere. All results are obtained by using a ray-velocity equation of motion derived from Fermat's principle.

**OCIS codes:** (080.0080) Geometric optics; (230.3205) Invisibility cloaks; (160.3918) Metamaterials; (080.2710) Inhomogeneous optical media.

**References and links**

1. J. B. Pendry, D. Schurig, and D. R. Smith, "Controlling Electromagnetic Fields," Science **312**, 1780-1782 (2006).
2. U. Leonhardt, "Optical Conformal Mapping," Science **312**, 1777-1780 (2006).
3. H. Chen, C.T. Chan, and P. Sheng, "Transformation optics and metamaterials," Nature Mater. 9, 387-396 (2010).
4. U. Leonhardt and T.G. Philbin, Geometry and Light: The Science of Invisibility (Dover, Mineola, 2010).
5. U. Leonhardt and T. Tyc, "Broadband Invisibility by Non-Euclidean Cloaking," Science **323**, 110 (2009).
6. J. Li and J. B. Pendry, "Hiding under the Carpet: A New Strategy for Cloaking," Phys. Rev. Lett. **101**, 203901 (2008).
7. R. Liu, C. Ji, J. J. Mock, J. Y. Chin, T. J. Cui, and D. R. Smith, "Broadband Ground-Plane Cloak," Science **323**, 366-369 (2009).
8. J. Valentine, J. Li, T. Zentgraf, G. Bartal, and X. Zhang, "An optical cloak made of dielectrics," Nature Mater. **8**, 568-571 (2009).
9. L. H. Gabrielli, J. Cardenas, C. B. Poitras, and M. Lipson, "Silicon nanostructure cloak operating at optical frequencies," Nature Photon. **3**, 461-463 (2009).
10. J. H. Lee, J. Blair, V. A. Tamma, Q. Wu, S. J. Rhee, C. J. Summers, and W. Park, "Direct visualization of optical frequency invisibility cloak based on silicon nanorod array," Opt. Express **17**, 12922-12928 (2009).
11. T. Ergin, N. Stenger, P. Brenner, J. B. Pendry, and M. Wegener, "Three-Dimensional Invisibility Cloak at Optical Wavelengths," Science **328**, 337-339 (2010).
12. H. F. Ma and T. J. Cui, "Three-dimensional broadband ground-plane cloak made of metamaterials," Nature Communications **1**, 1-6 (2010).
13. B. Zhang, Y. Luo, X. Liu, and G. Barbastathis, "Macroscopic Invisibility Cloak for Visible Light," Phys. Rev. Lett. **106**, 033901 (2011).

14. X. Chen, Y. Luo, J. Zhang, K. Jiang, J. B. Pendry, and S. Zhang, "Macroscopic invisibility cloaking of visible light," Nat. Commun. **2**, 176 (2011).
15. J. Fischer, T. Ergin, and M. Wegener, "Three-dimensional polarization-independent visible-frequency carpet invisibility cloak," Opt. Lett. **36**, 2059-2061 (2011).
16. M. Gharghi, C. Gladden, T. Zentgraf, Y. Liu, X. Yin, J. Valentine, and X. Zhang, "A Carpet Cloak for Visible Light,", Nano Lett. **11**, 2825–2828 (2011).
17. T. Ergin, J. Fischer, and M. Wegener, "Optical phase cloaking of 700-nm light waves in the far field by a three-dimensional carpet cloak," Phys. Rev. Lett., accepted for publication; arXiv:1107.4288v1.
18. R. Schmied, J. C. Halimeh, and M. Wegener, "Conformal carpet and grating cloaks," Opt. Express **18**, 24361-24367 (2010).
19. J. C. Halimeh, R. Schmied, and M. Wegener, "Newtonian photorealistic ray tracing of grating cloaks and correlation-function-based cloaking-quality assessment," Opt. Express **19**, 6078-6092 (2011).
20. D. Schurig. J. B. Pendry, and D. R. Smith, "Calculation of material properties and ray tracing in transformation media," Opt. Express **14**, 9794-9804 (2006).
21. A. Akbarzadeh and A. J. Danner, "Generalization of ray tracing in a linear inhomogeneous anisotropic medium: a coordinate-free approach," J. Opt. Soc. Am. A **27**, 2558-2562 (2010).
22. L. D. Landau, E. M. Lifshitz, and L. P. Pitaevskii, Electrodynamics of Continuous Media, Vol. 8 (Butterworth-Heinemann, Oxford, 1984).
23. M. Born and E. Wolf, Principles of Optics, 7. Ed. (University Press, Cambridge, 1999).
24. A.J. Danner, "Visualizing invisibility: Metamaterials-based optical devices in natural environments," Opt. Express **18**, 3332-3337 (2010).

## 1. Introduction

Transformation optics connects geometry of curved space and propagation of light in inhomogeneous anisotropic optical media *via* mapping physical path length onto optical path length [1-4]. This concept has, for example, provided various blueprints for macroscopic invisibility-cloaking structures in free space [1-5] as well as for the simplified carpet-cloak (or ground-plane) geometry [6]. Regarding the latter, several experiments have meanwhile successfully been performed ranging from microwave to visible frequencies [7-16].

Recently, early experiments have even introduced phase-sensitive imaging of the carpet cloak in the visible [17]. Optical phase is proportional to optical path length and proportional to the time of flight (TOF) of a light ray. These quantities are at the heart of transformation optics. Furthermore, measuring TOF images *via* LIDAR (light detection and ranging) is a well-established technology that could actually be used to reveal objects that may appear barely visible at first sight in mere amplitude imaging of some scenery.

In this paper, we calculate TOF images for three different types of invisibility cloaks, namely Gaussian and grating carpet cloaks [6,18] as well as of the paradigmatic cylindrical free-space cloak [1]. We also consider the invisible sphere (or 360-degree lens). The carpet cloaks and the invisible sphere are based on locally isotropic dielectrics, whereas the cylindrical cloak requires locally anisotropic magneto-dielectric yet impedance-matched effective materials. As the TOF is directly proportional to the optical path length in Fermat's principle, we chose to derive the ray equation of motion directly from Fermat's principle in this paper.

## 2. Carpet Cloak

Calculating two-dimensional TOF images requires knowledge of optical ray paths and refractive-index distribution along those paths in a three-dimensional optically inhomogeneous environment. The carpet cloak [6] employs locally isotropic refractive indices for which we have previously re-derived the ray equation of motion directly from Fermat's principle [19]. At the boundaries of the cloak, we use Snell's law and the Fresnel coefficients. On this basis, it is straightforward to numerically calculate the optical path length by integrating the refractive index along the ray path (Fermat's integral). The absolute time of flight results from the optical path length divided by the vacuum speed of light.

To eliminate trivial geometric contributions to the TOF images and to allow for a direct and intuitive assessment of cloaking/invisibility quality, all TOF images depicted in this paper are actually relative images, *i.e.*, they are the difference between the time-of-flight image with cloak and object and that without object and without cloak. The TOF images for the invisible sphere in section 5 are defined likewise. For a perfect cloak (or invisible object), this relative TOF should obviously be zero seconds for all image points.

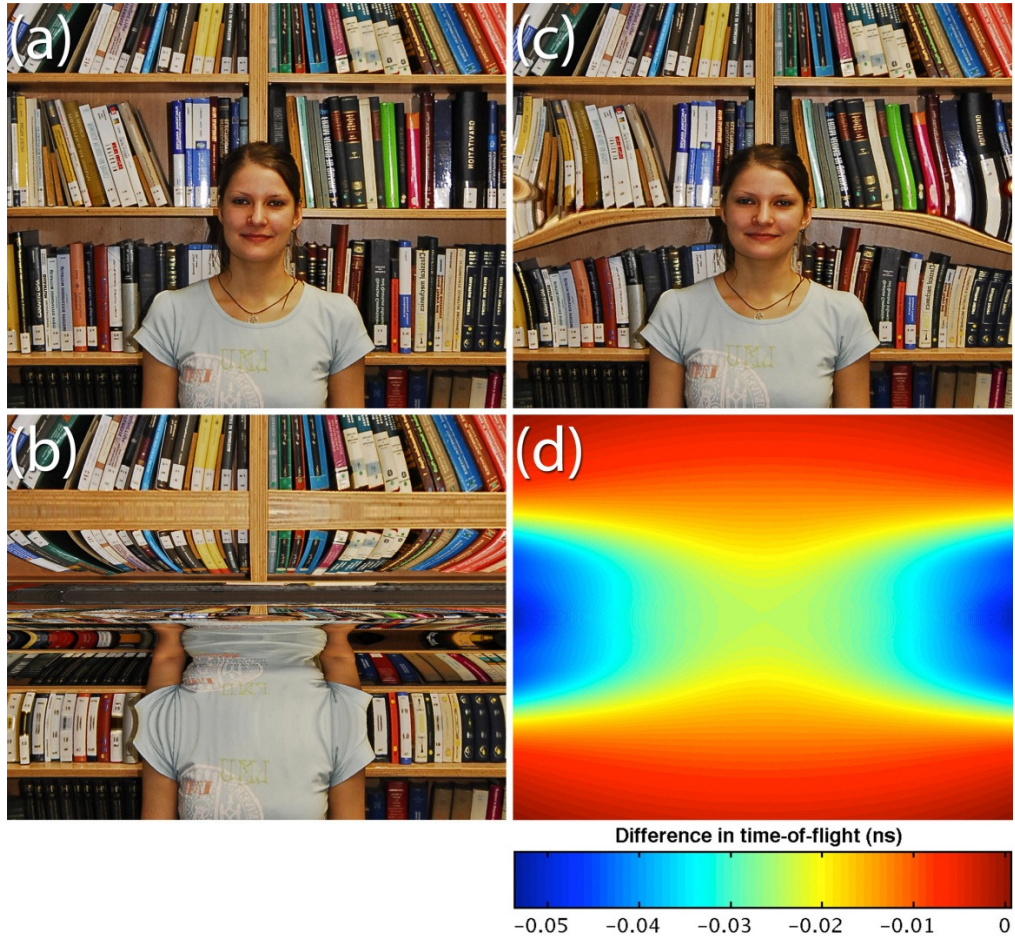

Fig. 1. Ray-tracing results for the Gaussian carpet cloak [6,18,19]. The scenery has previously been introduced [19]: a model is standing in front of a book shelf and is looking at her reflection in a mirror. (a) Rendered image without Gaussian bump in the mirror. (b) Rendered image with bump. (c) Rendered image with bump and with cloak. (d) Relative time-of-flight (TOF) image corresponding to the pixel-wise time delay between (c) and (a). The parameters correspond to those that we have previously used in Fig. 4 of Ref. 19. All ray-tracing calculations shown in this paper neglect the effects of dispersion. Thus, the color in panel (c) is merely for illustration.

We again use the conformal version of the carpet cloak [18] because it allows for obtaining an exact analytical refractive-index distribution (in contrast to the original quasi-conformal carpet cloak [6], which requires numerical evaluation). We discuss the Gaussian and the grating carpet cloaks [18,19] and start with the Gaussian version. To directly connect to our previous work, all parameters in this brief section are the same as in Ref. 19 (in particular, see Fig. 4 therein). Figure 1 (a)-(c) of the present work reproduces the previously published amplitude images, whereas panel (d) exhibits the numerically computed difference between the times of flight between panels (c) and (a). Obviously, large relative TOF deviations occur towards the edges of the image. These deviations are partly due to the fact that we (ab)use an originally two-dimensional design [6,18] in three dimensions. As expected, cloaking is very good in the middle of the image in Fig. 1. We have also computed TOF images for the conformal grating cloak [18,19] that are shown in Fig. 2. Even though the grating cloak exhibits bigger maxima, summing magnitudes over all pixels gives a lower total magnitude of the relative TOF than the Gaussian cloak by a factor of 1.52, indicating better cloaking performance in agreement with Ref. 19.

The bottom line of this brief section is that, for the carpet cloak, the distortions in the amplitude image show a clear correlation with large values in the TOF images. Alternatively, we can say that the amplitude distortions result from the fact that the optical path lengths are not quite right.

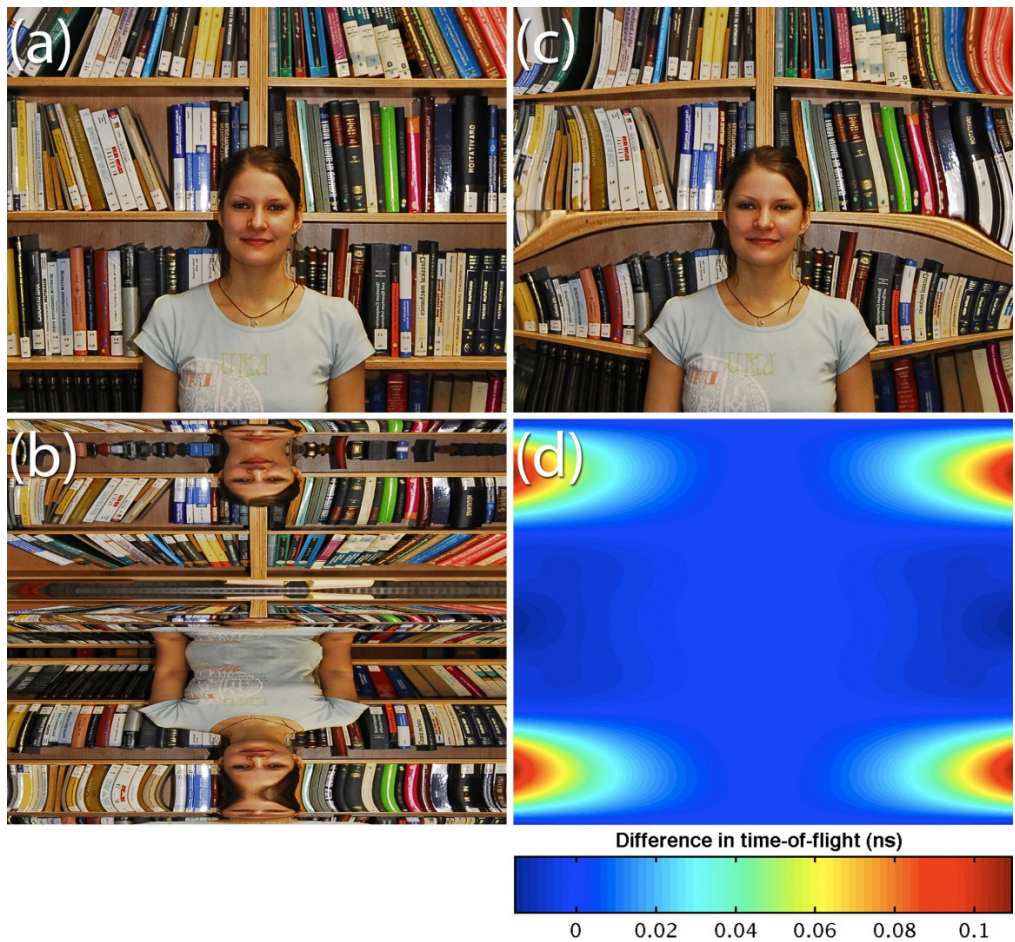

Fig. 2. As Fig. 1 but for the grating cloak [18,19]. (a) Rendered image without grating in the mirror. (b) Rendered image with grating. (c) Rendered image with grating and with cloak. (d) Relative time-of-flight (TOF) image corresponding to the pixel-wise time delay between (c) and (a). The parameters correspond to those that we have previously used in Fig. 3 of Ref. 19.

## 3. Lagrangian Ray Tracing

It is interesting to compare the findings of the previous section with other cloaks. Beyond the simple carpet cloak, generally locally anisotropic magneto-dielectric materials are required. Transformation optics always leads to impedance-matched media [1], for which the mathematics of Hamiltonian ray tracing has previously been developed [20]. Even the generalization beyond impedance-matched anisotropic magneto-dielectric (hence generally birefringent) structures has recently been published [21].

We here use a different approach for the case of impedance-matched media that is driven by two aims. First, we want to directly connect to Fermat's principle, which is analogous to the least-action principle in mechanics. Second, we wish to derive an equation of motion for the ray-velocity vector $\vec{v}$, which has an immediate and intuitive interpretation, namely the local direction and the magnitude of the velocity of light. In contrast, Ref. 20 obtained an equation of motion for the wave vector of light $\vec{k}$, which, in general, has no trivial connection to $\vec{v}$. Eventually, however, the two different approaches must deliver equivalent physical results, especially identical rendered amplitude and TOF images.

Using Fermat's principle in the first place implies that losses need to be negligible. Due to causality, this is equivalent to saying that the frequency dependence of the permittivity $\overleftrightarrow{\varepsilon}$ and the permeability $\overleftrightarrow{\mu}$ tensors needs to be negligible. In this case, we can use the following simple (electrostatic) expression for the electromagnetic energy density $w$

$$w = \frac{1}{2}(\vec{D} \cdot \vec{E} + \vec{B} \cdot \vec{H}) = \frac{1}{2}\left(\varepsilon_0(\overleftrightarrow{\varepsilon}\vec{E}) \cdot \vec{E} + \mu_0(\overleftrightarrow{\mu}\vec{H}) \cdot \vec{H}\right)$$

with the electric-field vector $\vec{E}$, the electric-displacement vector $\vec{D}$, the magnetic-field vector $\vec{H}$, the magnetic-induction vector $\vec{B}$, and the free-space permittivity $\varepsilon_0$ and permeability $\mu_0$. By using this expression for the electromagnetic energy density, we also tacitly imply that the components of the permittivity and permeability tensors are not negative. This means that, *e.g.*, effective metallic responses are excluded (*within* the transformation-optics structure), because they would lead to a negative, hence unphysical electromagnetic energy density. Impedance matching to vacuum means that the permittivity and permeability tensors are equal at each position in space, *i.e.*,

$$\overleftrightarrow{\varepsilon}(\vec{r}) = \overleftrightarrow{\mu}(\vec{r})$$

(for the case of impedance-matching to some other medium, the two can be proportional). All tensor components are real and the tensors are symmetric.

Fermat's principle postulates a maximal/minimal optical path length or TOF. We start with Fermat's principle in the general form [22]

$$\delta \int_{t_1}^{t_2} \vec{k} \cdot \vec{v} \, dt = 0$$

where $\vec{v}$ is the energy-velocity or ray-velocity vector and $\vec{k}$ is the wave vector of light. Using the Maxwell equations, the energy-velocity vector can be expressed as [23]

$$\vec{v} = \frac{\vec{S}}{w} = \frac{\vec{E} \times \vec{H}}{w}$$

where $\vec{S}$ is the Poynting vector. The wave vector of light can be expressed as

$$\vec{k} = \frac{\omega}{w}\vec{D} \times \vec{B}$$

where $\omega$ is the angular frequency of light. Inserting this expression into Fermat's principle, we obtain

$$\delta \int_{t_1}^{t_2} \frac{\omega}{w} \left( \vec{D} \times \vec{B} \right) \cdot \vec{v} \, dt = 0$$

$$= \delta \int_{t_1}^{t_2} \frac{\omega}{w} \left( \left( \varepsilon_0 \overleftrightarrow{\varepsilon} \vec{E} \right) \times \left( \mu_0 \overleftrightarrow{\varepsilon} \vec{H} \right) \right) \cdot \vec{v} \, dt$$

Using the mathematical identity

$$\left( \overleftrightarrow{A} \vec{a} \right) \times \left( \overleftrightarrow{A} \vec{b} \right) = |\overleftrightarrow{A}| \, \overleftrightarrow{A}^{-1} \left( \vec{a} \times \vec{b} \right)$$

and dividing by the constants, we get

$$\delta \int_{t_1}^{t_2} \left( |\overleftrightarrow{\varepsilon}| \, \overleftrightarrow{\varepsilon}^{-1} \left( \frac{\vec{E} \times \vec{H}}{w} \right) \right) \cdot \vec{v} \, dt = \int_{t_1}^{t_2} \left( |\overleftrightarrow{\varepsilon}| \, \overleftrightarrow{\varepsilon}^{-1} \vec{v} \right) \cdot \vec{v} \, dt = 0$$

where |...| denotes the determinant of the tensor. Introducing the auxiliary matrix

$$\overleftrightarrow{M}(\vec{r}) = |\overleftrightarrow{\varepsilon}| \, \overleftrightarrow{\varepsilon}^{-1}$$

Fermat's principle can finally be cast into the compact form

$$\delta \int_{t_1}^{t_2} \vec{v} \cdot \left( \overleftrightarrow{M}(\vec{r}) \, \vec{v} \right) dt = 0$$

$$=: \delta \int_{t_1}^{t_2} L(\vec{r}, \vec{v}, t) \, dt$$

In the last step, we have highlighted the analogy to Hamilton's principle in classical mechanics based on the Lagrangian $L$ (also see Ref. 19). In perfect mathematical analogy to the Euler-Lagrange equations in classical mechanics

$$\frac{\partial L}{\partial r_i} - \frac{d}{dt} \frac{\partial L}{\partial v_i} = 0$$

leading to Newton's second law, we derive the ray equation of motion (*i.e.*, the light-ray acceleration)

$$\frac{d\vec{v}}{dt} = \frac{\overleftrightarrow{M}^{-1}}{2} \left( \vec{v} (\vec{\nabla} \overleftrightarrow{M}) \vec{v} - 2 \left( (\vec{\nabla} \overleftrightarrow{M}) \cdot \vec{v} \right) \vec{v} \right)$$

In this compact notation, $\vec{\nabla} \overleftrightarrow{M}$ is a vector, the components of which are tensors, namely the partial derivatives of the auxiliary matrix $\overleftrightarrow{M}(\vec{r}) = |\overleftrightarrow{\varepsilon}| \, \overleftrightarrow{\varepsilon}^{-1} = |\overleftrightarrow{\mu}| \, \overleftrightarrow{\mu}^{-1}$ with respect to the three spatial coordinates $r_i$ ($i = 1,2,3$).

Within the cylindrical cloaks under investigation, these ordinary differential equations can, *e.g.*, be solved numerically by using a fourth-order Runge-Kutta method. We note in passing that we have programmed our approach as well as that of Ref. 20 and that we find better convergence for ours. At the boundaries of the cloak, where the optical properties may change discontinuously, we use Snell's law and the Fresnel coefficients. This means that we do account for reflections, which, however, are so weak that they are not visible in the below rendered images. This is to be expected as the cylindrical cloak (unlike the carpet cloak above) is, by design, impedance-matched to vacuum at its boundaries.

## 4. Cylindrical Free-Space Cloak

The scenery for the free-space cloaks in this section is the same as in section 2. However, for the free-space cloaks, a virtual point camera looks at the model standing in front of a book shelf (see Fig. 3(a)). The corresponding raw images used for this view as well as for the view into the opposite direction in the room are shown in Fig. 3(b) and (c). The object to be made invisible is a perfectly reflective metal cylinder with radius $a=5$ cm. In cylinder coordinates, the magneto-dielectric parameters of the cylindrical cloak with outer radius $b=10$ cm around that metal cylinder are given by the radial ($r$), azimuthal ($\Theta$), and axial components ($z$) [1]

$$\varepsilon_r = \mu_r = \frac{r-a}{r}; \quad \varepsilon_\Theta = \mu_\Theta = \frac{r}{r-a}; \quad \varepsilon_z = \mu_z = \left(\frac{b}{b-a}\right)^2 \frac{r-a}{r}$$

Ray-tracing results are shown in Fig. 4. The representation corresponds to that in Figs. 1 and 2 for the Gaussian and the grating carpet cloaks, allowing for direct comparison. Obviously, cloaking in Fig. 4 is perfect to within our numerical accuracy (note the time scale in (d)). Indeed, we can also view this result as a successful demanding control computer-experiment for the achieved numerical accuracy of our ray-tracing approach because the cylindrical cloak is mathematically expected to be perfect under these conditions – even in three dimensions, for any viewing angle, and for any polarization of light [1]. After all, in sharp contrast to the carpet cloak [6], the cylindrical cloak [1] is based on an exact three-dimensional transformation-optics result.

The price one pays is that the ideal cylindrical cloak requires some optical tensor components approaching zero and others approaching infinity at radius $r \to a$ within the cloak. Thus, it is interesting to study the effect of truncation, which will likely occur in some form or the other in any real-world experiment. To mimic and visualize such limitations by example, we *ad hoc* set all permittivity and permeability tensor components <0.1 to a constant value of 0.1 and likewise tensor components >10 to a constant value of 10. The resulting cloaking behavior is depicted in Fig. 5. Here, distortions occur in the amplitude as well as in the relative TOF images of the cloak.

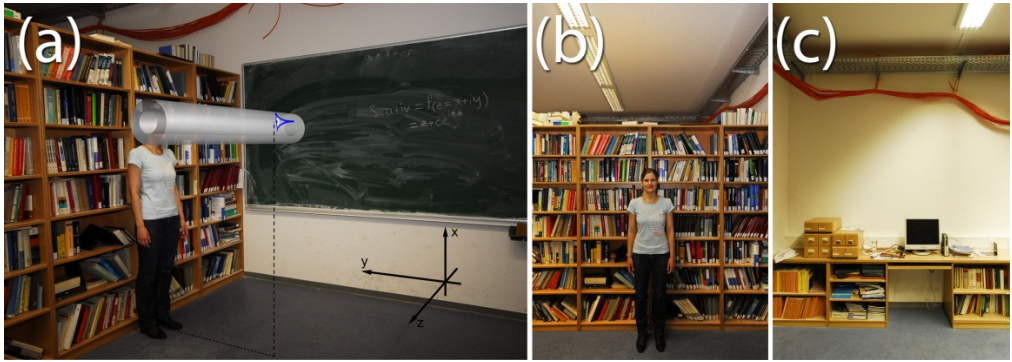

Fig. 3. (a) Illustration of the scenery used for the ray-tracing calculations for all free-space invisibilities in Figs. 4-6. A virtual point camera (the "eye") looks at a model standing in front of a book shelf. The point camera is at a distance of 100 cm from the middle of the model's eyes and at the same height above the floor with a field of view (FOV) of 42° vertically and 50° horizontally, emulating a human focal FOV. The axis of the metal cylinder and of the concentric cylindrical cloak in Figs. 4 and 5 as well as the center of the invisible sphere in Fig. 6 are at a distance of 45.5 cm from the middle between the model's eyes and at the same height above the floor. (b) and (c) depict the raw images used as input for the ray-tracing calculations.

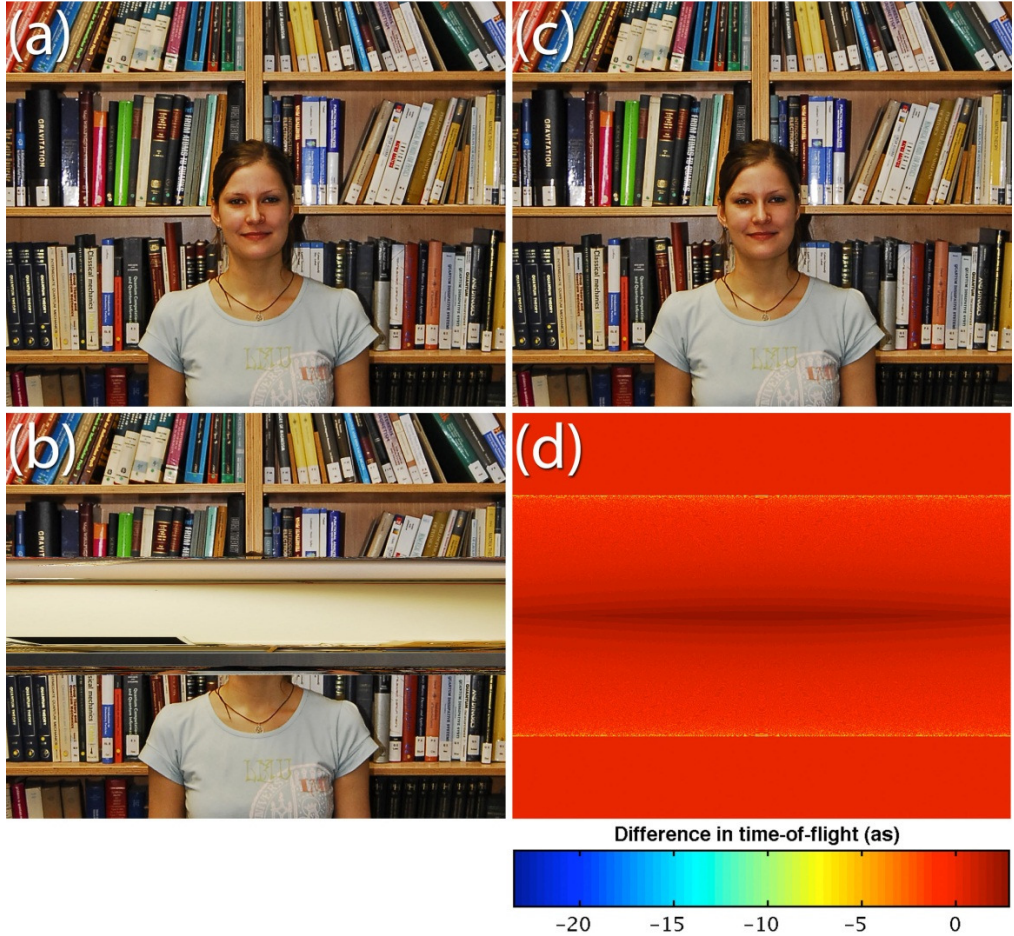

Fig. 4. Ray-tracing results for the exact cylindrical free-space cloak [1]. In this scenery, which is illustrated in Fig. 3, a virtual point-camera looks at a model standing in front of a book shelf. (a) Rendered image without metal cylinder (compare Fig. 1(a)). (b) Rendered image with horizontal metal cylinder in front of model. Reflections from this cylinder show the other side of the room. (c) Rendered image with metal cylinder and with cylindrical cloak. (d) Relative time-of-flight (TOF) image corresponding to the pixel-wise time delay between (c) and (a). Note the units of attoseconds (1 as = $10^{-18}$ s) here rather than nanoseconds (1 ns = $10^{-9}$ s) as in Figs. 1(d), 2(d), 5(d), and 6(c).

## 5. The Invisible Sphere

In general, the TOF images provide information independent from the amplitude images. To emphasize and to exemplify this point, we finally consider the so-called invisible sphere [24] (or 360-degree lens). This device is *not* an invisibility cloak. It is just a dielectric sphere that ideally has no effect on the amplitude image at all. It is supposed to look as if nothing was there. Thus, this example provides another easy-to-interpret control computer-experiment for our numerical ray-tracing approach. In contrast, we expect a pronounced effect in the TOF images as the rays circle around inside the sphere. Hence, they accumulate substantial additional optical path length or, equivalently, additional TOF.

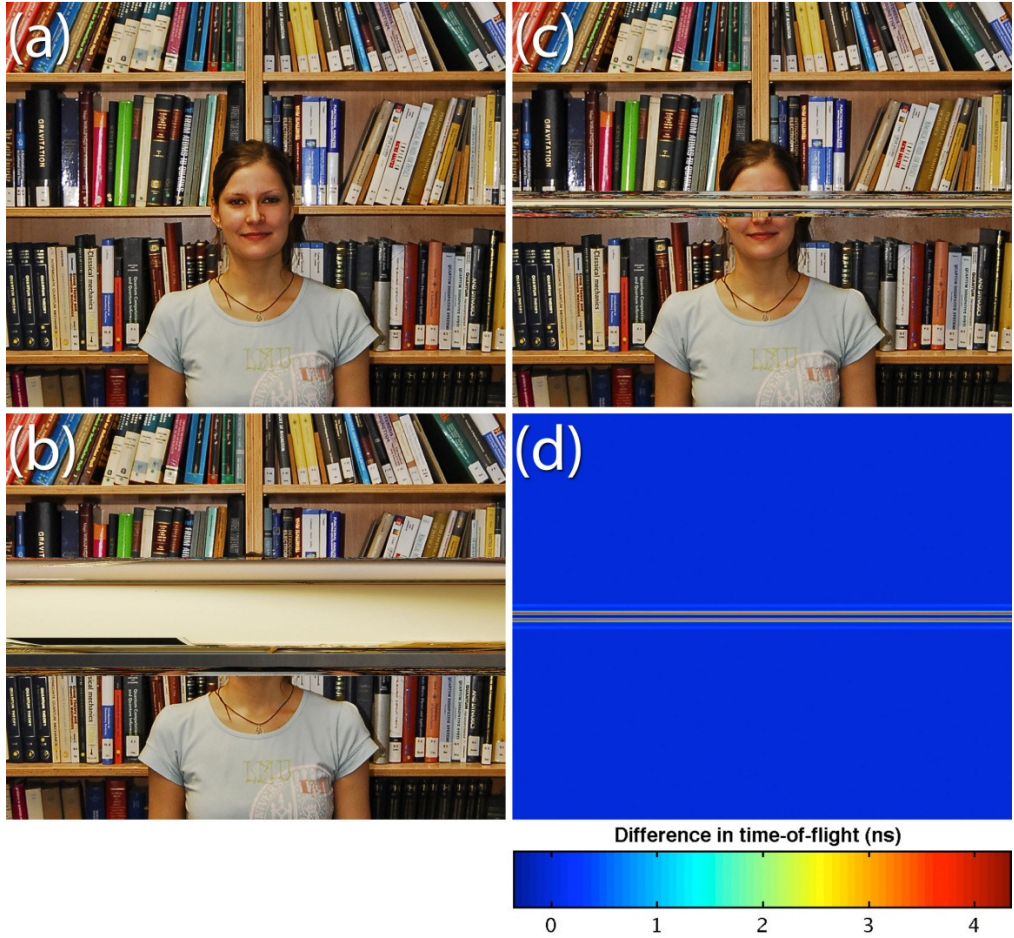

Fig. 5. Ray-tracing results for the truncated cylindrical free-space cloak (compare with ideal cylindrical free-space cloak shown in Fig. 4). Here, tensor components below 0.1 are set to 0.1 and tensor components larger than 10 are set to 10.

Precisely, the invisible sphere [24] is a sphere of radius $R$ filled with a locally isotropic dielectric with a radial dependence of the spherically symmetric refractive-index profile $n(r)$ given by the implicit form

$$\sqrt{n} = \frac{R}{r\,n} + \sqrt{\left(\frac{R}{r\,n}\right)^2 - 1}$$

In the center of the sphere, we have $n(0) = \infty$, at its surface $n(R) = 1$. Thus, the sphere is even impedance-matched to the surrounding air/vacuum, leading to vanishing reflections too. As $n(r) \geq 1\ \forall\ r \leq R$, all rays entering the sphere accumulate additional optical path length within the sphere compared to free space.

In the corresponding numerical ray-tracing calculations shown in Fig. 6, we have used a sphere radius of $R$=10 cm and the same scenery as for the cylindrical cloaks in the previous section. The "invisible sphere" is indeed barely visible in Fig. 6(b), evidencing excellent numerical convergence *via* our approach [19], despite the presence of the refractive-index singularity in the center of the sphere. In sharp contrast, the "invisible sphere" has a dramatic effect on the relative TOF image in Fig. 6(c). Our calculations show that the difference in the TOF of the invisible sphere and that of empty space is constant and given by the circumference of the sphere $2\pi R$ divided by the vacuum speed of light. After all, the "invisible sphere" is not really invisible if one picks the right observable.

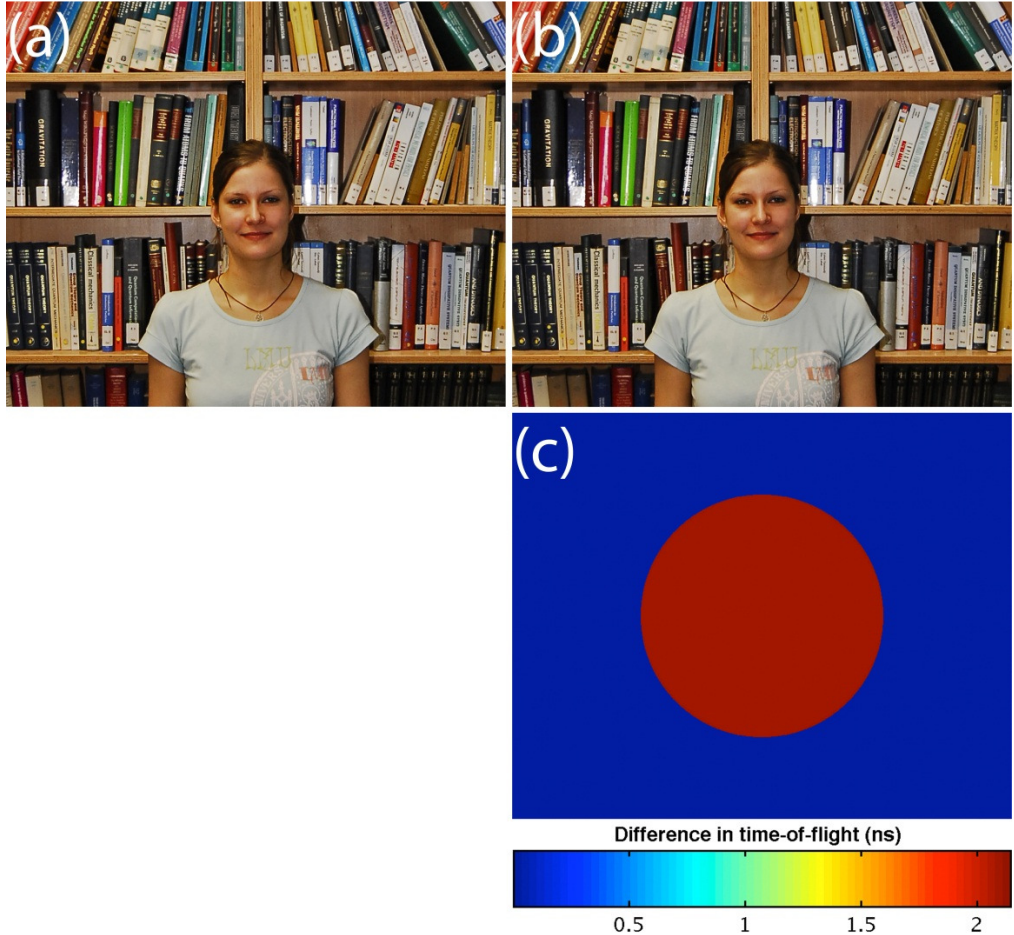

Fig. 6. Ray-tracing results for the invisible sphere. The scenery is as in Figs. 4 and 5. (a) Rendered image without the sphere. (b) Rendered image with invisible sphere. (c) Relative time-of-flight (TOF) image corresponding to the pixel-wise time delay between (b) and (a).

## 6. Conclusion

We have numerically calculated time-of-flight (TOF) images of the carpet cloak, the grating cloak, the cylindrical free-space cloak, and the invisible sphere by ray tracing using an equation of motion for the ray-velocity vector directly derived from Fermat's principle. For both the Gaussian and the grating conformal carpet cloaks, the TOF deviations correlate with the distortions in the rendered amplitude images. The grating cloak outperforms the Gaussian carpet cloak. The cylindrical cloak is strictly perfect, but truncation of its singular parameters leads to imperfect amplitude and TOF images as well. For the invisible sphere, the amplitude images ideally show zero effect, whereas a strong TOF effect is expected and found. Thus, broadly speaking, TOF images (*e.g.*, experimentally obtained by LIDAR) provide an interesting additional check of the quality of cloaking/invisibility and could be used to uncover hidden objects. Indeed, recent visible-frequency, far-field, phase-sensitive imaging experiments on the three-dimensional carpet cloak [17] make first steps in this direction.

**Acknowledgements**

We thank Oleg Yevtushenko, Immanuel Bloch, and Tanja Rosentreter (LMU München) as well as Roman Schmied (Universität Basel) for discussions. We also thank our model, Tanja Rosentreter, and the photographer, Vincent Sprenger (TU München), for help with the photographs used as input for the photorealistic ray-tracing calculations. We are grateful to Maximilian Papp and Husni Habal (TU München) for help with Fig. 3(a) and the aesthetic arrangement of the figures. We acknowledge the support of the Arnold Sommerfeld Center (LMU München), which allowed us to use their computer facilities for our rather CPU-time-consuming numerical ray-tracing calculations. J.C.H. acknowledges financial support by the Excellence Cluster "Nanosystems Inititiative Munich (NIM)". M.W. acknowledges support by the Deutsche Forschungsgemein-schaft (DFG), the State of Baden-Württemberg, and the Karlsruhe Institute of Technology (KIT) through the DFG Center for Functional Nanostructures (CFN) within subproject A1.5. The project PHOME acknowledges the financial support of the Future and Emerging Technologies (FET) programme within the Seventh Framework Programme for Research of the European Commission, under FET-Open grant number 213390. The project METAMAT is supported by the Bundesministerium für Bildung und Forschung (BMBF).